# Low temperature phase transitions and crystal structure of $Na_{0.5}CoO_2$

Q. Huang[1], M.L. Foo[2], J.W. Lynn[1], H.W. Zandbergen[3], G. Lawes[4], Yayu Wang[5], B. H. Toby[1], A.P. Ramirez[6], N.P. Ong[5], and R.J. Cava[2]

[1]Center for Neutron Research, NIST, Gaithersburg MD, USA; [2]Department of Chemistry, Princeton University, Princeton NJ, USA; [3]National Centre for High Resolution Electron Microscopy, Technical University Delft, The Netherlands; [4]Department of Thermal Physics, Los Alamos National Laboratory, Los Alamos, NM, USA; [5]Department of Physics, Princeton University, Princeton NJ, USA; [6]Lucent Technologies Bell Laboratories, Murray Hill NJ, USA.

**Abstract**

The crystal structure of $Na_{0.5}CoO_2$, determined by powder neutron diffraction, is reported. The structure consists of layers of edge-shared $CoO_6$ octahedra in a triangular lattice, with Na ions occupying ordered positions in the interleaving planes. The Na ions form one-dimensional zigzag chains. Two types of Co ions, which differ only subtly in their coordination by oxygen, are also found in chains. Specific heat measurements show that the transitions observed at 87 K and 53 K in the resistivity and magnetic susceptibility are accompanied by changes in entropy. Electron diffraction studies suggest that the 87K transition may have a structural component.

**Introduction**

Investigations of the copper oxide superconductors have revealed much about the unexpected behavior of charge transport in square lattices. Though triangle-based lattices have been studied in electrically insulating geometrically frustrated magnets for some time (1-3), conductive layered triangular lattices have not emerged for detailed study until recently, with the discovery of a surprisingly high thermopower in the hexagonal symmetry metallic conductor $Na_{0.7}CoO_2$ (4). This compound consists of triangular layers of cobalt between close-packed layers of oxygen, resulting in an array of edge shared $CoO_6$ octahedra, separated by layers of Na that partially fill sites in the hexagonal plane between the layers of $CoO_6$ octahedra. The large thermopower has been shown to be due to the transport of spin entropy (5). Interest was further stimulated in this compound by the discovery of superconductivity at 4.5 K in $Na_{0.3}CoO_2·1.3H_2O$ (6) at a different level of electronic band filling than gives rise to the high thermopower in the same cobalt oxide lattice in $Na_{0.7}CoO_2$.

In order to study the transition between the high thermopower state and the host state for the superconductivity, we recently completed a transport-based study (7) of the electronic phase diagram for single crystals of non-hydrated $Na_xCoO_2$ for $0.3<x<0.75$. The data revealed a crossover in properties from an unusual Curie-Weiss metal near $x=0.7$ to a paramagnetic metallic state for x=0.3. Surprisingly, a distinct compound at $Na_{0.5}CoO_2$, which has a transition to an insulating state at 53 K, separates the two metallic regimes at higher and lower sodium concentrations. Characterization of $Na_{0.5}CoO_2$ by electron diffraction revealed the presence of an orthorhombic symmetry superlattice, which was attributed to Na ordering and an associated charge ordering of the underlying Co. Here we report specific heat measurements on $Na_{0.5}CoO_2$ to characterize further the low temperature phase transitions, and electron diffraction patterns showing the presence of a structural distortion at low temperature. In addition, the average crystal structure of $Na_{0.5}CoO_2$, determined at 300 K and 3.5 K by neutron powder diffraction, is reported. Two types of Co chains, associated with zigzag chains of Na running along one crystallographic direction, are found.

**Experimental**

Compounds with the γ-$Na_xCoO_2$ structure type, the host phase for the superconductor and the thermoelectric, can be synthesized only in the range of stoichiometry 0.7<x<0.8 by conventional thermodynamic methods. Lower x contents of the same structure type were synthesized by chemical deintercalation of Na at room temperature. The $Na_{0.5}CoO_2$ powder employed here for the neutron diffraction, specific heat, and susceptibility measurements was synthesized from single phase powder of $Na_{0.75}CoO_2$ prepared by heating appropriate proportions of $Na_2CO_3$ and $Co_3O_4$ overnight in flowing oxygen at 800 C. That powder was then treated by immersion (with stirring) in acetonitrile saturated with $I_2$ in sufficient excess to insure full oxidation of the powder to $Na_{0.5}CoO_2$. The powder was then washed with acetonitrile and stored in a dry environment for all subsequent handling. Small single crystals of $Na_{0.5}CoO_2$ were obtained by the same method by starting with floating-zone grown crystals of $Na_{0.7}CoO_2$. Further details have been described elsewhere (7).

Magnetization measurements were carried out between 300 and 1.8 K in an applied field of 1 kOe, employing a Quantum Design MPMS magnetometer. Standard four-contact resistivity measurements were carried out on small single crystals (approximately 1 mm x 2 mm x 0.1 mm) between 300 K and 2 K. Specific heat was measured between 2 K and 150 K on a polycrystalline powder of total mass 0.1 g, mixed with silver powder to insure thermal contact, in a Quantum Design PPMS system. The specific heat of the silver powder and the addenda were measured separately and subtracted from the raw data for $Na_{0.5}CoO_2$. Electron diffraction measurements were performed in Philips CM300UT and CM200ST electron microscopes, both equipped with a field emission gun. Nanodiffraction was performed using a condenser aperture of 10 μm and an electron probe size of 10-30 nm in diameter. The specimen cooling experiments were performed using a custom-made holder for operation at about 100 K.

The neutron powder diffraction intensity data for $Na_{0.5}CoO_2$ were collected using the BT-1 high-resolution powder diffractometer at the NIST Center for Neutron Research, employing a Cu (311) monochromator to produce a neutron beam of wavelength 1.5403(1) Å. Collimators with horizontal divergences of 15′, 20′, and 7′ arc were used before and after the monochromator, and after the sample, respectively. The intensities were measured in steps of 0.05° in the 2θ range 3°-168°. Data were collected at a variety of temperatures from 300 K to 3.5 K to elucidate the possible magnetic and crystal structure transitions. The structural parameters were refined using the program GSAS (8). The neutron scattering amplitudes used in the refinements were 0.363, 0.253, and 0.581 ($\times 10^{-12}$ cm) for Na, Co, and O, respectively.

**Results**

Figure 1 shows the resistive and magnetic characterization of the low temperature phase transitions in $Na_{0.5}CoO_2$. The resistivity measured in the $CoO_2$ plane increases significantly on cooling at an onset temperature of approximately 53 K. After a slight saturation, the slope of the ρ vs. T curve increases with cooling below 25 K, indicating another phase transition. The magnetic susceptibility, on the other hand, shows decreases at 87 K and 53 K, but no features at 20 K. The 87 K phase transition can be observed in the resistivity data, but only by taking the derivative of the curve (inset, fig. 1). Characterization of $Na_{0.5}CoO_2$ by μSR has indicated the onset of magnetic ordering at 53 K, resolving further into the ordering of two distinct types of magnetic atoms at 20 K. No magnetic ordering was associated with the 87 K transition (9).

Specific heat characterization of the phase transitions is shown in figures 2 and 3. For the 87 K transition, a distinct peak is observed, characteristic of a three-dimensional phase transition. For the transition at 53 K, the specific heat change is more subtle, and appears to be somewhat broader in temperature than the 87 K transition. Figure 3 shows in the main panel that there is no specific heat anomaly for the 20 K transition, indicating no substantial change in magnetic entropy. This suggests that this may be a spin reorientation transition. The data further show the presence of a small peak in the specific heat at 3K. The integrated entropy in this transition is very small, approximately that expected for 0.1% or less of the ordering of spin ½ particles. To check whether this could be intrinsic to the phase, we performed low temperature susceptibility measurements, shown in the inset. No feature is seen in the susceptibility, suggesting that this may indeed be due to the presence of an impurity phase.

Figure 4 compares the room temperature and low temperature [001] (basal plane) electron diffraction patterns for $Na_{0.5}CoO_2$. The cooling stage is expected to reach temperatures on the order of 100 ± 20 K. The ambient temperature electron diffraction pattern (Figure 4(a)) shows the orthorhombic a√3,2a superstructure, described in detail below. Extra diffraction spots are clearly seen in the low temperature diffraction pattern (Figure 4(b)) indicative of a structural distortion at low temperatures in which both the *a* and *b* axes of the orthorhombic cell are tripled. These spots are not observed in all crystallites in the sample. This suggests that such super-super cell ordering may be susceptible to damage by the electron beam, or exist over only a narrow range of temperature, or that compositional inhomogeneities make it favorable in some areas but not in others. The intensity of the extra spots is substantially weaker than those of the hexagonal substructure spots and the orthorhombic superstructure spots, indicating the subtlety of the structural distortion. This distortion may be associated with the 87 K phase transition observed in other characterization experiments, but we have not unambiguously determined whether that is the case. In conjunction with the other data, these data suggest that this transition may be associated with the formation of a charge density wave or orbital ordering. Further study will be required to characterize this super-superlattice ordering in more detail.

A structural model for $Na_{0.5}CoO_2$ with hexagonal $P6_3/mmc$ symmetry and lattice parameters a≈2.81 and c≈11.2 Å was first used in the structural refinement employing the neutron diffraction data. The model gives a good fit to all the strongest intensity peaks, with the Na atoms distributed in a disordered fashion in the 2$b$ and 2$c$ sites within the $P6_3/mmc$ symmetry space group with the occupancy parameter of approximately 0.25 for both sites. The data, however, show a substantial number of additional weak Bragg peaks at all measured temperatures (Fig. 5). These peaks can be indexed by the orthorhombic superlattice with $a_o=\sqrt{3}\ a_H$, $b_o=2a_H$, $c_o = c_H$, where $a_H$ and $c_H$ are hexagonal sublattice parameters, originally observed by electron diffraction (7). A superstructure model having $Pnmm$ symmetry and the Na atoms completely ordered (Figs. 6 and 7) was derived based on semi-quantitative analysis of electron diffraction data, and was tested quantitatively with the neutron diffraction data. The model provides a very good fit to the neutron diffraction data, and unambiguously represents the average crystal structure of $Na_{0.5}CoO_2$. The refined structural parameters and selected bond distances and angles are shown in Table 1. Fig. 5(b) and fig. 8 show a good fit of the calculated intensities to the observed neutron powder diffraction pattern at 295 K. We note that our instrumental resolution is insufficient to resolve any metric deviations from the ideal $a_o=\sqrt{3}\ a_H$, and $b_o=2a_H$ relationships, and therefore the orthorhombic $a$ and $b$ axes were not varied independently in the structural refinements. Also, due to the hexagonal pseudosymmetry, some of the positional parameters for some atoms were constrained to be fixed at ideal positions for the final refinements. There is no indication of significant deviations from these ideal positions. The average structure is not significantly different at 3.5 K. Though the average structure is well described in the present analysis, elucidation of finer details of the crystal structure will have to await higher resolution studies.

The crystal structure of $Na_{0.5}CoO_2$ is unexpected. In the simplest case, if Na positions are determined by ion-ion repulsion within the Na layer, then a triangular Na lattice is expected for x = 1/3, but x = 1/2 is not expected to yield a special case of long range ordering. In $Na_xCoO_2$, however, the availability of two types of Na sites that have comparable energies adds an extra degree of freedom to the possible structures. Both are trigonal prismatic sites, but one trigonal prism shares edges with adjacent $CoO_6$ octahedra (Na2), while the other trigonal prism shares faces with adjacent $CoO_6$ octahedra (Na1). The Na2 site is preferred in approximately a 2:1 ratio for a broad range of Na contents in $Na_xCoO_2$ (8), but at $Na_{0.5}CoO_2$ we find the Na1 and Na2 sites are occupied in equal ratios.

Figures 9 (a) and (b) show the structure from different perspectives. The first is based on a central Co plane,

and the Na layers above and below it, and shows the zigzag chains of Na and the parallel chains of Co over several unit cells. The second extracts the Na atoms in the ordered layer only. This figure shows that the Na lattice is in fact a distorted hexagonal lattice. To make the lattice into an ideal hexagonal arrangement, which would minimize the $Na^+$-$Na^+$ Coulomb repulsions, the Na ions would be forced into unfavorably shaped Na-O coordination polyhedra. Thus the geometry the Na displays must involve a compromise between ion-ion repulsion and optimally shaped Na-O coordination polyhedra

This kind of compromise cannot explain, however, why an ordered structure would form at all at this composition – with Na ions actually moving into sites (Na1) at x = 0.5, given that these sites are not favored at other compositions due to the Na-Co repulsion across the shared triangular oxygen face. We must conclude that x = 0.5 is a special composition for electronic reasons, as is borne out by its unique magnetic and electronic properties. The crystal structure determination reinforces this interpretation because two distinct types of Co atoms are seen, with the parallel chain configuration already described. We postulate that the Na then "decorates" the particular Co chains in which the Co oxidation state is lower. Thus the final Na configuration is a three-way compromise: accommodating the effects of interaction with the underlying Co lattice (whose charge configuration we believe to be determined by a fundamental electronic instability), the mutual repulsion of the Na ions in the layer, and the locations of Na-O coordination polyhedra of favorable shape.

Finally, figure 10 highlights three lines of $CoO_6$ octahedra within a single layer of Co. The figure is drawn to emphasize the number and distribution of long and short Co-O bonds for each of the two types of Co chains. The figure shows that Co1 has one short and 5 long bonds to O, whereas Co2 has two short and 4 long bonds to O. This suggests that the Co1 is lower in oxidation state than Co2. The Na in the planes below and above the Co plane, while bonded to O in the coordination polyhedra of both Co1 and Co2, are coordinated to a larger fraction of the O in the $(Co1)O_6$ polyhedron. This supports our conjecture that the Na chooses to decorate the chain in which the less highly charged Co is present, consistent with Pauling's rules. The distribution of long and short bonds suggests the possibility of orbital ordering within the chains. The differences in observed bond lengths are very subtle. Therefore structure refinements using higher precision diffraction data will be needed to allow an unambiguous determination of the Co charge or orbital ordering in this compound.

**Conclusion**
The $Na_xCoO_2$ phase diagram naturally divides into two metallic and magnetic regimes, one at high Na content that exhibits unusual spin transport properties, and the other at low Na content that is the host composition for the superconductor. The composition that divides these regimes, $Na_{0.5}CoO_2$, undergoes several phase transitions below 100 K, and becomes a magnetically ordered insulator in the ground state. The crystal structure involves a novel ordering of the Na ions into zigzag chains that we postulate decorate chains of Co ions with different oxidation states. Temperature dependent higher precision diffraction studies would be of great interest to determine further details of the crystal structure and any structural changes that occur at the three phase transitions. The ordered structure that this compound exhibits should facilitate further electronic band structure calculations, and comparison to those for other compositions would be of great interest to help clarify why $Na_{0.5}CoO_2$ exhibits such novel electronic, structural, and magnetic properties.
.

**Acknowledgements**
The research at Princeton University was supported by the NSF MRSEC program, grant DMR-0213706 and the NSF Solid State Chemistry Program grant DMR 0244254. Identification of commercial equipment in the text is not intended to imply recommendation or endorsement by the National Institute of Standards and Technology

**References**
1. A.P. Ramirez, in Handbook of Magnetic Materials. Ed. K. J. H. Buschow. New York: New Holland, (2001).
2. Greedan, J.E., *J. Mater. Chem.* **11,** 37-53 (2001)
3. Moessner, R., *Can. J. Phys.* **79,** 1283-94 (2001)
4. Terasaki, I., Sasago, Y. and Uchinokura, K., *Phys. Rev.* B**56**, R12685 – R12687 (1997).
5. Wang, Yayu, Rogado, N.S., Cava, R.J., and Ong, N.P., *Nature* **423** 425 (2003)
6. Takada, K., Sakurai, H., Takayama-Muromachi, E., Izumi, F., Dilanian, R.A., and Sasaki T., *Nature* **422**, 53 – 55 (2003)
7. Maw Lin Foo, Yayu Wang, Satoshi Watauchi, H. W. Zandbergen, Tao He, R. J. Cava, and N. P. Ong, cond-mat/0312174
8. Larson, A., and Von Dreele, R.B. Los Alamos National Laboratory, Internal Report (1994).
9. T. J. Uemura, P.L. Russo, A.T. Savici, C.R. Wiebe, G.J. MacDougall, G.M. Luke, M. Mochizuki, Y. Yanase, M. Ogata, M.L. Foo, R.J. Cava, to be published.

**Table I.** Structural parameters and selected bond distances (Å) and angles (degrees) for Na$_{0.5}$CoO$_2$ at 295 K (first line) and 3.5 K (second line). Space Group: P*nmm* (#59) [P*nmm* ***bca*** > P*mmn* (Origin choice 2)].

a= 4.87618(5)    b= 5.63053(9)    c= 11.1298(2) Å
   4.87570(4)       5.62997(7)       11.0634(2)

| Atom  | Site | x        | y   | z         | B (Å$^2$) |
|-------|------|----------|-----|-----------|-----------|
| Co(1) | 4f   | 0        | 1/4 | 0         | 0.49(3)   |
|       |      | 0        | 1/4 | 0         | 0.23(2)   |
| Co(2) | 4d   | 1/2      | 0   | 0         | 0.49(3)   |
|       |      | 1/2      | 0   | 0         | 0.23(2)   |
| Na(1) | 2b   | -0.027(2)| 1/4 | 3/4       | 1.7(1)    |
|       |      | -0.018(2)| 1/4 | 3/4       | 0.79(6)   |
| Na(2) | 2a   | 0.361(2) | 3/4 | 3/4       | 1.7(1)    |
|       |      | 0.352(2) | 3/4 | 3/4       | 0.79(6)   |
| O(1)  | 4f   | 1/3*     | 1/4 | 0.0894(1) | 0.51(1)   |
|       |      | 1/3*     | 1/4 | 0.0895(1) | 0.29(1)   |
| O(2)  | 4f   | 1/3*     | 3/4 | 0.0816(3) | 0.51(1)   |
|       |      | 1/3*     | 3/4 | 0.0820(3) | 0.29(5)   |
| O(3)  | 8g   | -1/6*    | 0*  | 0.0894(1) | 0.51(1)   |
|       |      | -1/6*    | 0*  | 0.0895(1) | 0.29(1)   |

Selected bond distances and angles

| | | | | |
|---|---|---|---|---|
| Co(1)-O(1) |     | 1.9059(6) | O(1)-Co(1)-O(3) | 95.22(4) |
|            |     | 1.9029(6) |                 | 95.41(4) |
| Co(1)-O(2) |     | 1.862(6)  | O(2)-Co(1)-O(3) | 96.76(5) |
|            |     | 1.862(2)  |                 | 96.85(4) |
| Co(1)-O(3) | ×4  | 1.9059(6) | O(3)-Co(1)-O(3) | 95.22(4) |
|            |     | 1.9029(6) |                 | 95.41(6) |
| Co(2)-O(1) | ×2  | 1.9059(3) | O(1)-Co(2)-O(2) | 96.76(5) |
|            |     | 1.9029(6) |                 | 96.85(4) |
| Co(2)-O(2) | ×2  | 1.862(3)  | O(1)-Co(2)-O(3) | 95.22(4) |
|            |     | 1.862(2)  |                 | 95.41(4) |
| Co(2)-O(3) | ×2  | 1.9059(6) | O(2)-Co(2)-O(3) | 96.76(4) |
|            |     | 1.9029(6) |                 | 96.85(4) |

| | | |
|---|---|---|
| Na(1)-O(2) | ×2 | 2.396(7) |
| | | 2.411(6) |
| Na(1)-O(3) | ×4 | 2.464(4) |
| | | 2.439(3) |
| Na(2)-O(1) | ×3 | 2.328(7) |
| | | 2.348(2) |
| Na(2)-O(3) | ×3 | 2.464(4) |
| | | 2.439(2) |

---

- $R_p$= 3.13 %　　$R_{wp}$= 3.85 %　　$\chi^2$= 1.250
  　　3.35　　　　　　　4.11　　　　　　1.288

Due to high correlations
- Structural parameters were constrained to be $a/b = \sqrt{3}/2$, and $z(O(1))=z(O(3))$.
- Temperature factors for Co, Na, and O were constrained to be equal, respectively.
- * Parameters fixed to hexagonal pseudosymmetry positions.

**Figure Captions**

**Fig 1.** The magnetic and resistive characterization of the phase transitions in $Na_{0.5}CoO_2$ below 100K. The resistive data are taken on a single crystal, measured in the $CoO_2$ plane. The magnetic susceptibility data are taken on the polycrystalline sample employed in the specific heat and neutron diffraction studies.

**Fig. 2.** Thermodynamic characterization of the 87K and 53K transitions in $Na_{0.5}CoO_2$.

**Fig. 3.** Specific heat in the vicinity of the 20 K transition. Inset, the magnetic susceptibility at low temperatures.

**Fig. 4.** [001] electron diffraction patterns of $Na_{0.5}CoO_2$. **(a)** The room temperature diffraction pattern of the orthorhombic $a\sqrt{3},2a$ superstructure. Orthorhombic cell indicated by the black rectangle. The 100 and 010 reflections of the basic hexagonal lattice are indicated. (Weak additional reflections are due to twinning.) **(b)** Diffraction pattern taken at about 80-100 K. The extra reflections require a tripling of both the *a* and *b* axes of the orthorhombic unit cell.

**Fig. 5.** Portion of the observed (crosses) and calculated (solid line) powder neutron diffraction intensities for $Na_{0.5}CoO_2$ at 295 K. **(a)** fitting using the sublattice with *P6$_3$/mmc* symmetry; **(b)** fitting by a superstructure model with *Pnmm* symmetry, where the peaks with index and shorter vertical lines are the superlattice peaks.

**Fig. 6.** The crystal structure of $Na_{0.5}CoO_2$, showing layers of edge-shared $CoO_6$ octahedra and the triangular prismatic coordination of Na within the intermediary layers.

**Fig. 7.** The Na ion configuration for $Na_{0.5}CoO_2$. The dotted lines show the *ab* plane of the hexagonal sublattice, and the solid lines indicate the orthorhombic superlattice.

**Fig. 8.** Observed (crosses) and calculated (solid line) intensities for $Na_{0.5}CoO_2$ at 295 K. Differences between the observed and calculated intensities are shown in bottom of figure. The vertical lines indicate the Bragg peak positions for $Na_{0.5}CoO_2$ (upper) and 1 wt% CoO (lower).

**Fig. 9.** **(a)** Shows a central layer of Co at z = ½ with a layer of Na above it (at z= ¾) and below it (at z = ¼) (see also figure 6). The zig-zag Na chain is emphasized as is the hexagonal Co lattice. Oxygen planes omitted for clarity. **(b)** The Na ion configuration in a single layer, over several unit cells, in $Na_{0.5}CoO_2$. The solid lines show an ideal hexagonal lattice.

**Fig. 10.** The arrangement of the longer and shorter Co-O bonds within the cell of $Na_{0.5}CoO_2$, with the adjacent Na layers shown. The designations of the Co and Na atoms are as indicated.

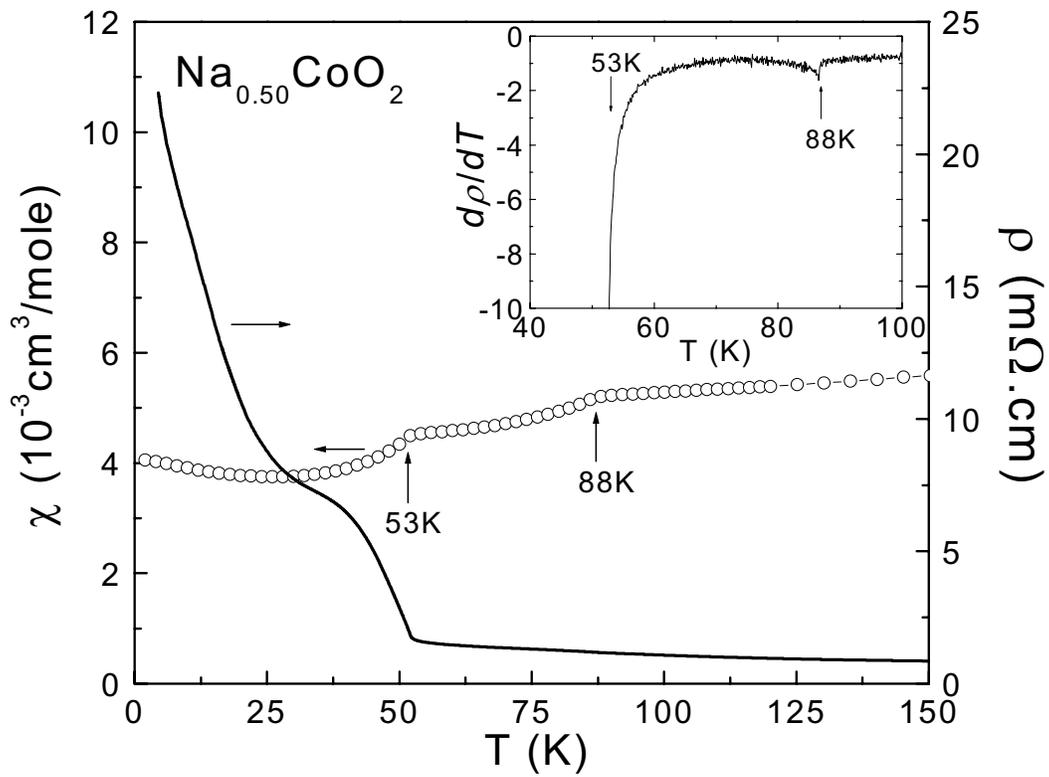

**Figure 1**

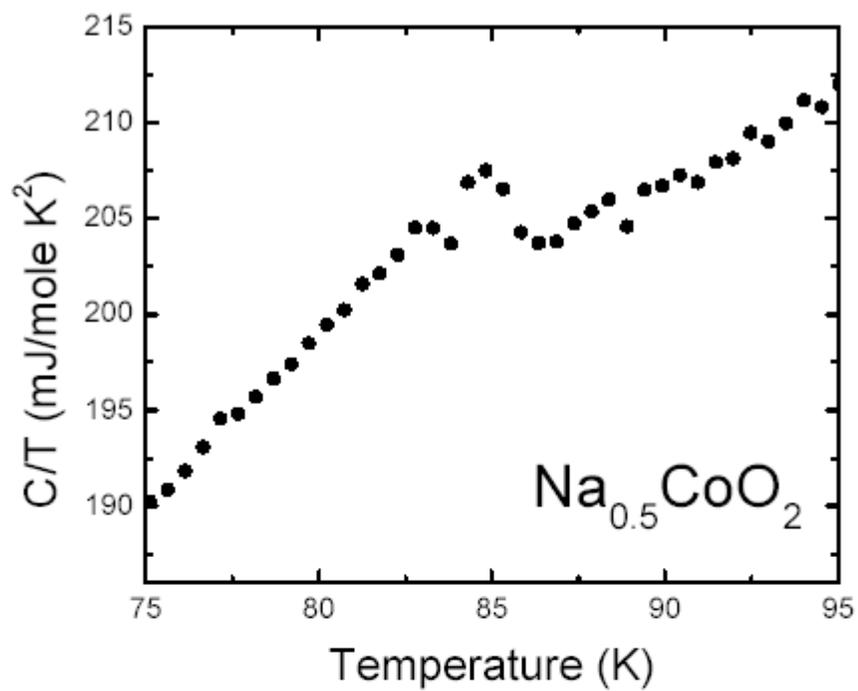

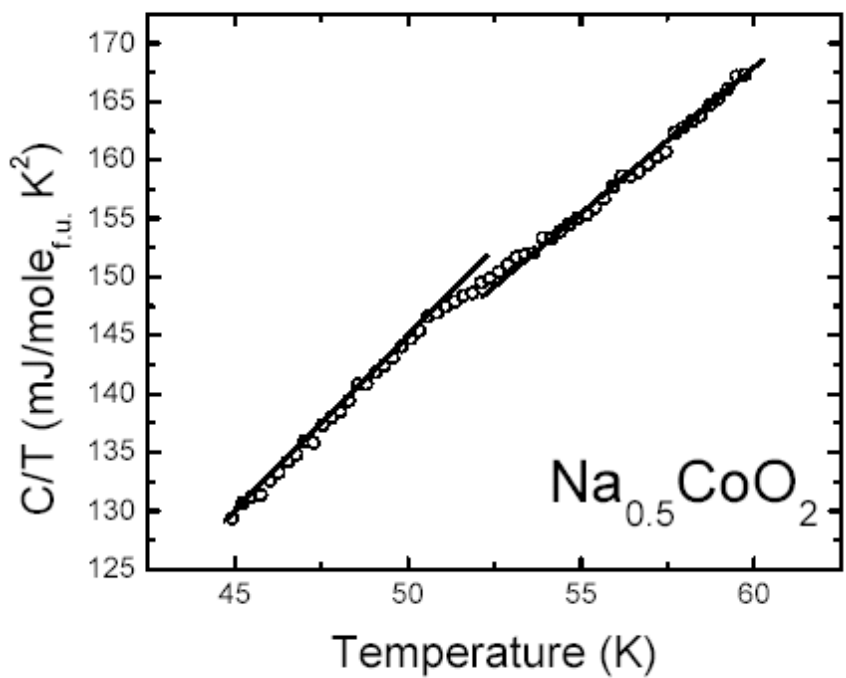

**Fig 2 a and b**

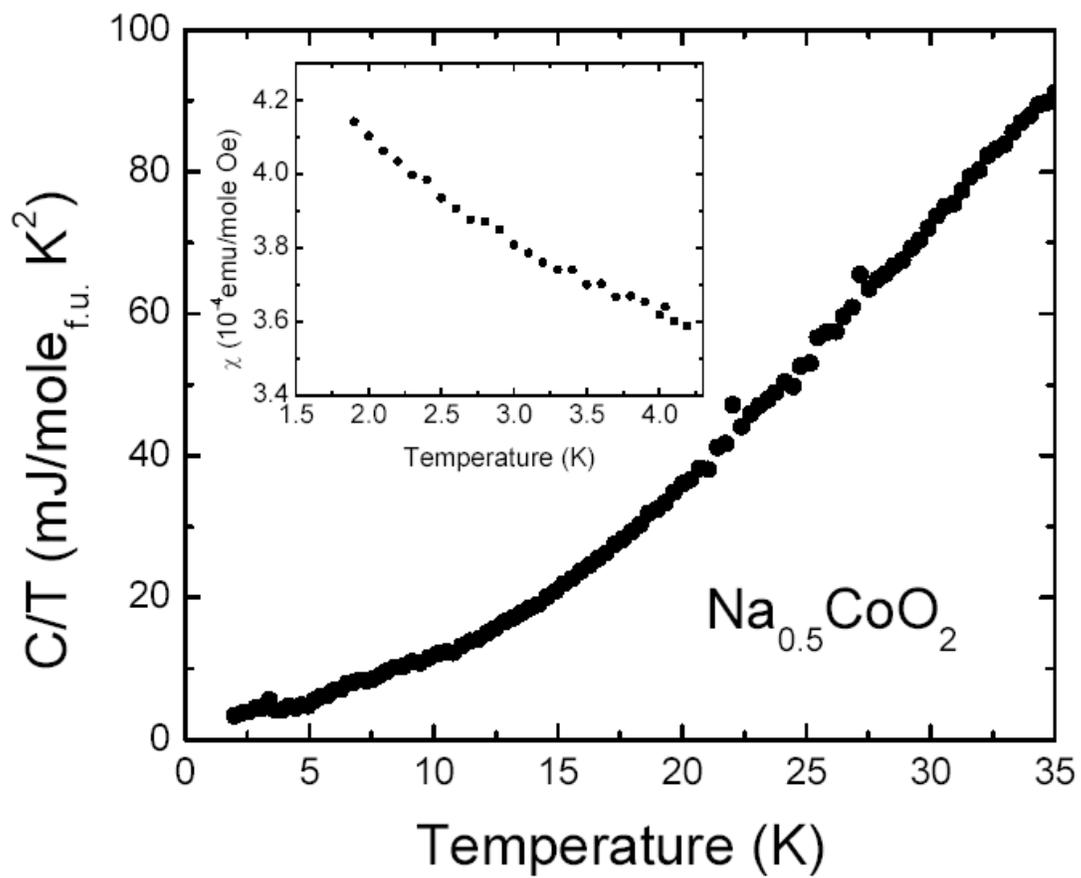

**Fig 3**

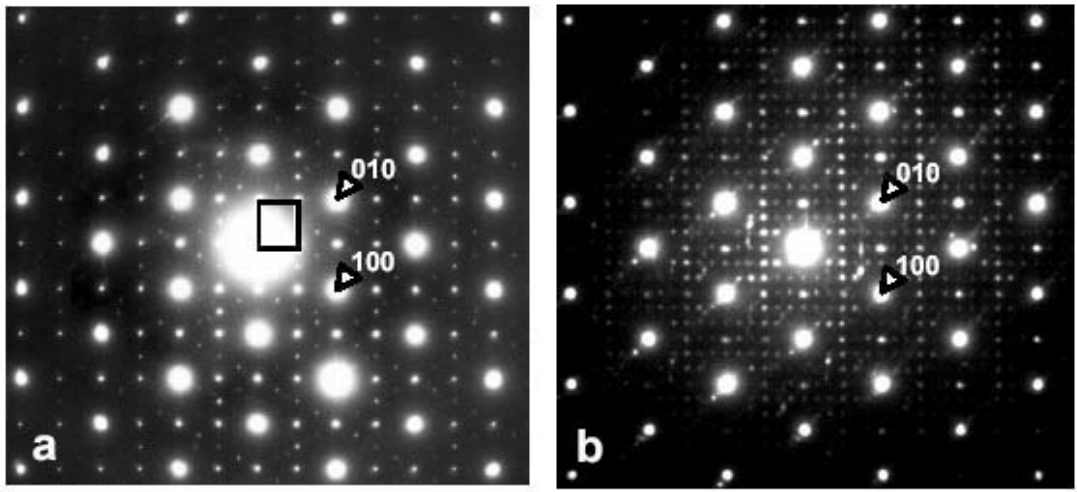

**Fig. 4**

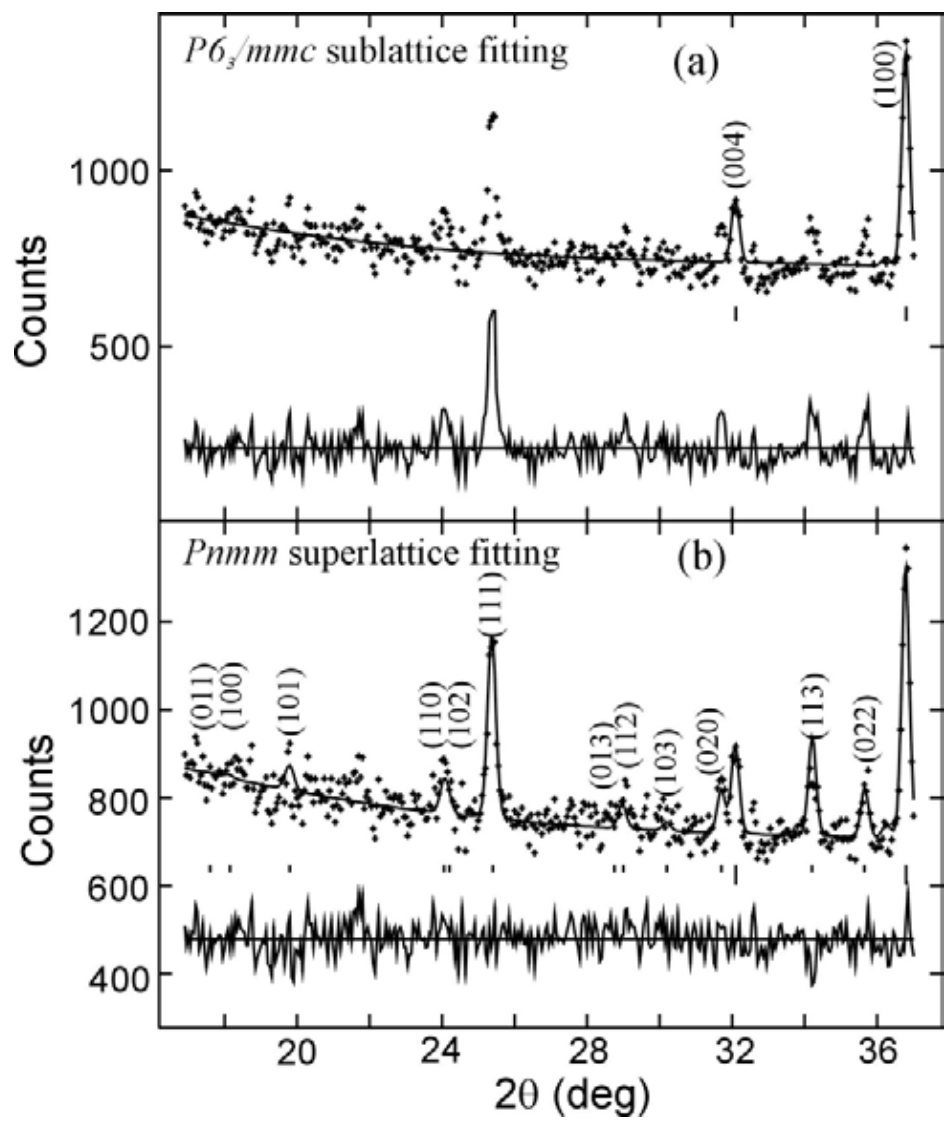

**Fig. 5**

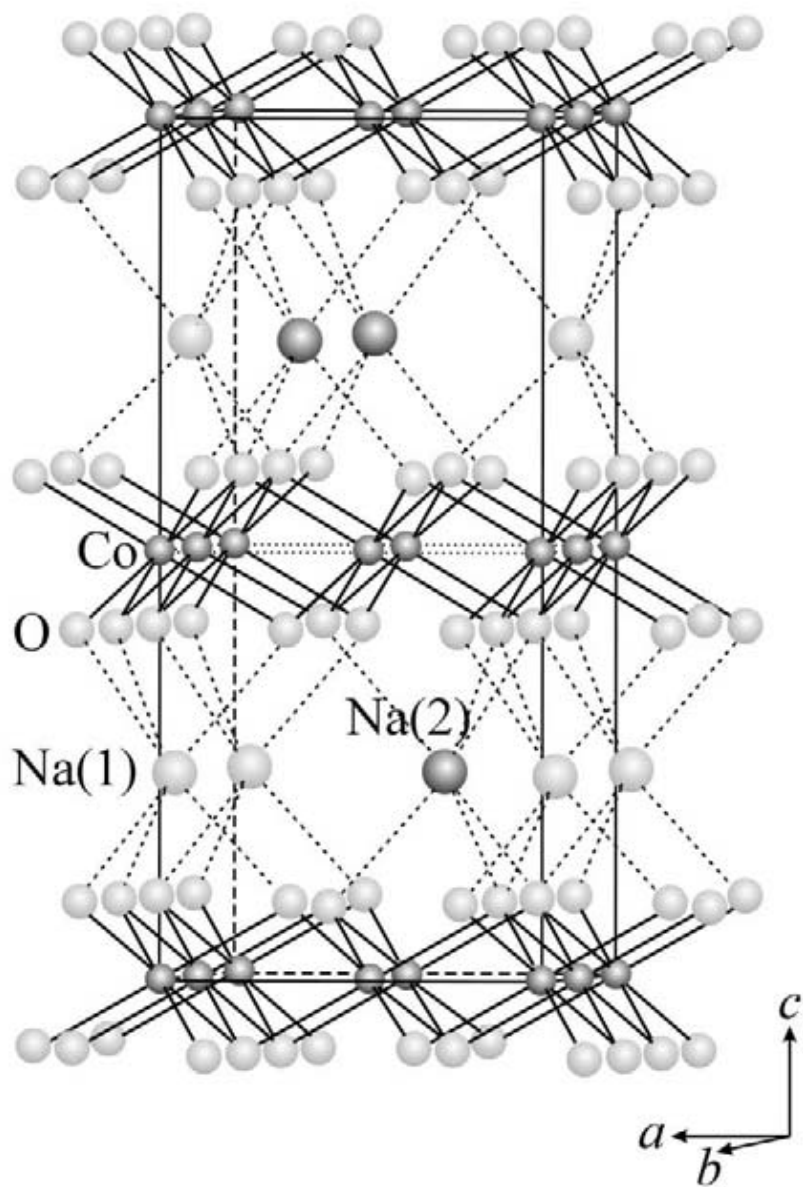

**Fig. 6**

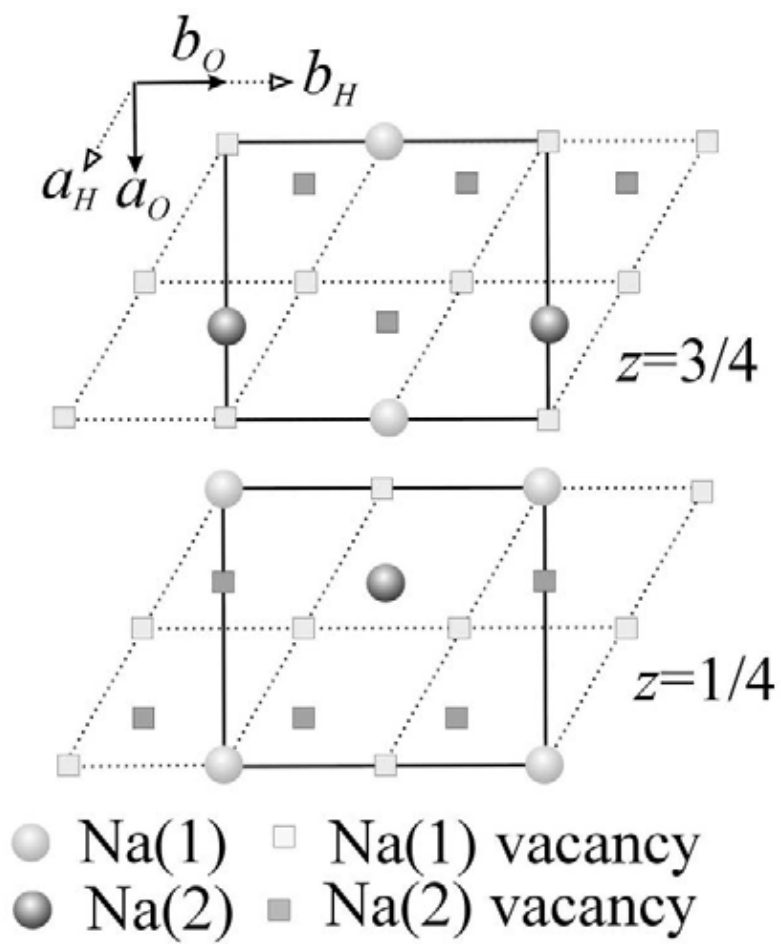

**Fig. 7**

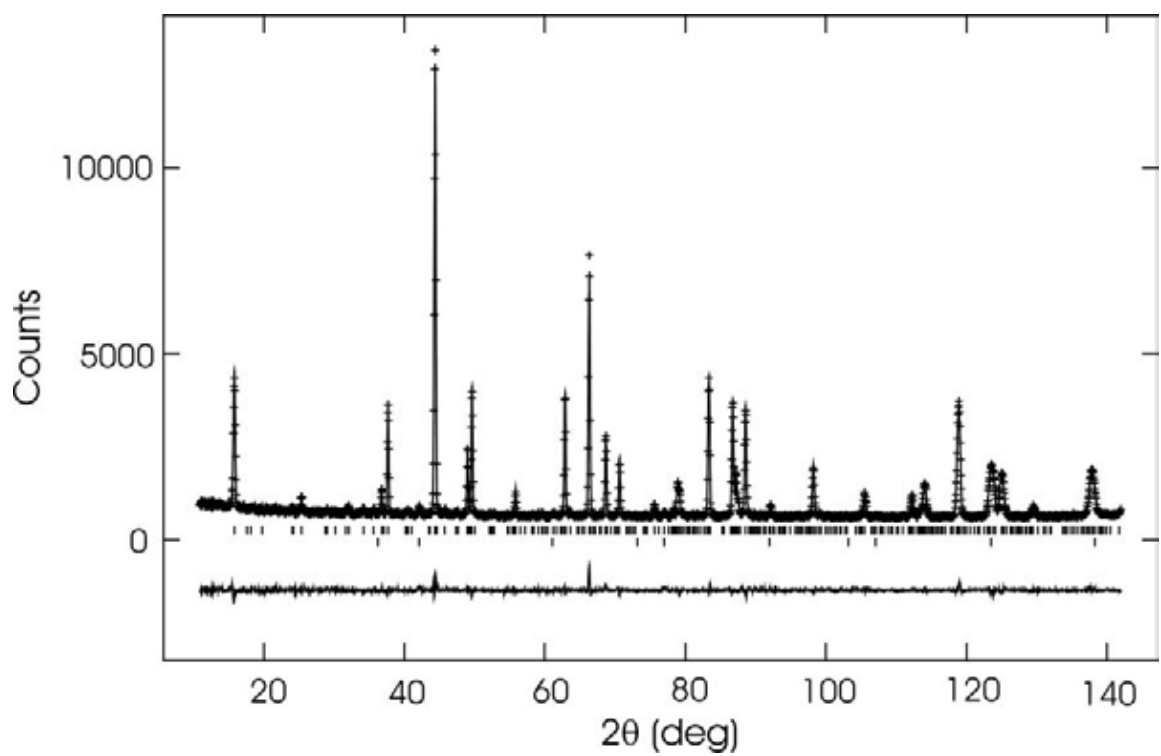

**Fig. 8**

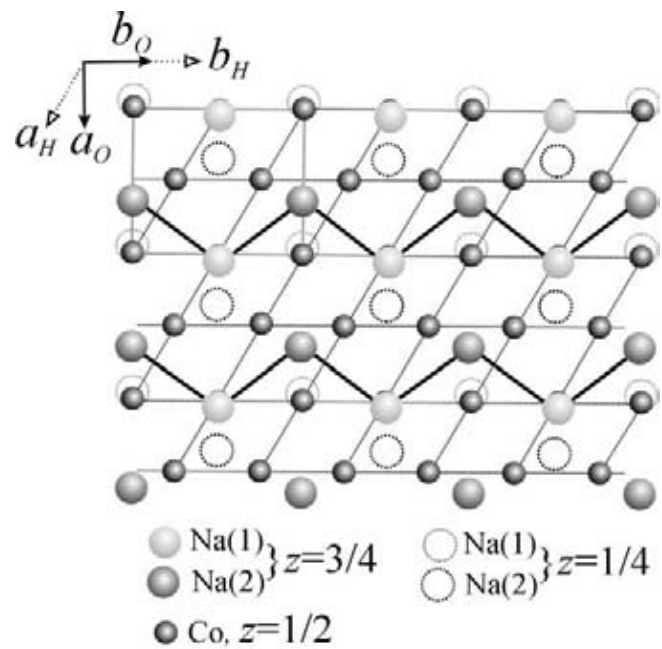

(a)

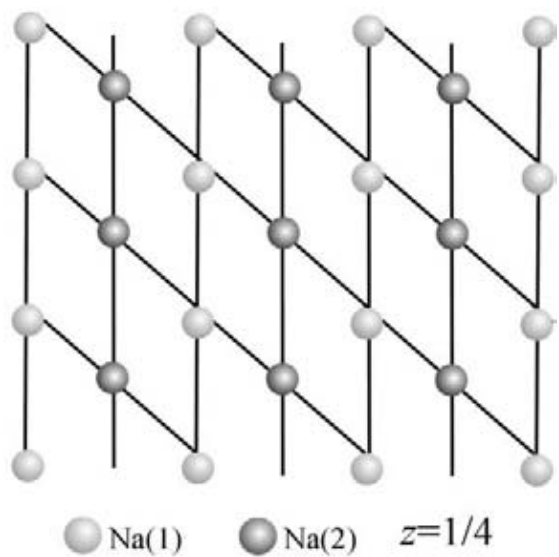

○ Na(1)  ● Na(2)   z=1/4

(b)

**Fig. 9**

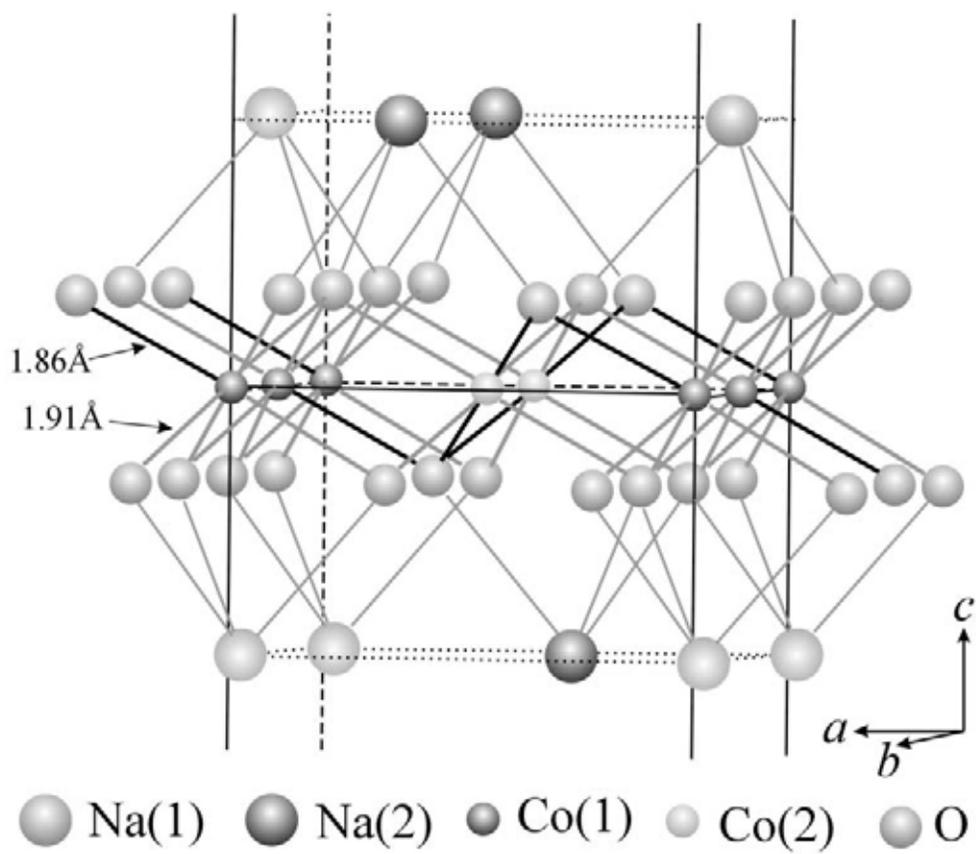

**Fig. 10**